# Giovanni de la Fontana, engineer and magician

**Amelia Carolina Sparavigna**
Department of Applied Science and Technology, Politecnico di Torino, Italy

*Giovanni de la Fontana was a man of the Italian Renaissance of the early fifteenth-century. Besides becoming a doctor of medicine, he was an engineer who used his skills to mimic the magicians, creating a sort of "artificial magic".*

The Italian Renaissance was an extraordinary period that produced a great cultural change in the Western Europe. It began in Italy during the 14th century and lasted until the 16th century, marking therefore the transition between the Medieval and Early Modern ages. The word Renaissance (Rinascimento in Italian) indicates a renewed interest in classical antiquities: however, this is a modern term that began to be used in the 19th century, after the work of Jules Michelet and Jacob Burckhardt.

During the Renaissance, artists and engineers changed the urban environment of Italian towns, enriching them of buildings and monuments. They even contributed to move the philosophy of nature towards a modern science. However, as acknowledged by several scholars [1,2], to these engineers, the occult sciences were parts of their worldview, such as to several other thinkers of the middle and early modern ages. These thinkers for instance, devoted themselves to the study of astrology that, in particular during the Renaissance, was used to approach some practical and intellectual problems. For instance, the astrologer Girolamo Cardano developed a work on medical astrology. Besides astrology, we find alchemy, a practice concerned with the technologies of minerals and metals and therefore a part of the natural philosophy, often used in medicine [2].

A recent publication is discussing the connection of technology with magic: in 2002, Anthony Grafton published the book entitled "Magic and technology in early modern Europe" [3]. In this book we find the developments of technology in the framework of the practice of magic. That is, according to Grafton, it was in an early modern Europe that "the notion of a mathematical or artificial magic rose from the ancient art of natural magic, embodied in such things as prayers and talismans, to the complex optical, hydraulic and mechanical devices". The machines and automata simulating the nature and movements of living beings "were so astonishing at first that the engineers had to seek independent authoritative testimony that the devices were the product of engineering skill, not occult practices." [3]

And this is why, before discussing the connection of engineering and magic, Grafton is proposing an episode where we find Athanasius Kircher, (1600–1680), German Jesuit well-known for his humanistic studies and researches in geology and medicine, involved in investigating the practice of a "magus". Shortly after 1650, a performer named Jean Royer appeared in the Piazza Navona at Rome [4]. He stood next to the fountain of the four rivers, just created by Gianlorenzo Bernini. Standing next to the marvelous mechanical fountain, Royer challenged it. People saw Royer first ingurgitate large quantities of water and then regurgitate it as whatever liquid the audience asked for vinegar, or wine. This performance provoked suspicion of witchcraft and demonic possession. However Royer found assistance from Kircher, and his friend Gaspar Schott. Schott observed the Royer's performance and brought him to meet Kircher: at the end Royer revealed his tricks to make water into vinegar or wine.

It was a time, during which clever tricks were easily misunderstood as magic or even witchcraft.

**A magic technology**
As told us by Ref.3, from ancient times until the early modern age, it was natural to believe that learned magi can practice the ancient art of natural magic. And so, using the powers of the heavens and "pulling on the subtle cosmic web of influence" down from the stars and planets of heavens to stones, plants, and the parts of human body, some thinkers devised talismans and spells. "These men, Kircher and Schott firmly believed, carried on a magic art that had its origins in ancient Egypt and that rested on true knowledge about the occult properties, the hidden sympathies and antipathies of all things in the universe. This same form of magic underpinned some of Kircher's own proudest inventions - like the sunflower clock, which turned with the sun, supposedly thanks to a universal magnetic force that kept it moving." [3] However, at the same time, there were some magi who practiced a second art, called by Kircher and Schott, the "artificial magic", because they used optical, hydraulic and mechanical techniques, rather than magical influences, to create their mechanical devices. The masters of this kind of magic were engineers.

"In the fifteenth century, however, both the status of magic and the claims of technology underwent metamorphoses. - Reference 3 continues - Clerics unprovided with benefices and learned men steeped in the magical traditions of the Islamic and Byzantine worlds began to write positively about magic, in a number of different forms. Traditional forms of magic ... offered their practitioners rapid and easy ways to do everything ... More sophisticated forms of magic, based on ancient disciplines like astrology and the Cabala, enabled scholars at a higher level to offer their customers luxurious new regimens that would keep them healthy, enhancing their life force and protecting them from malevolent planets like Saturn. The greatest intellects proclaimed that natural magic ... was not a dangerous indulgence in traffic with devils but a profound and legitimate discipline". And then Giovanni Pico della Mirandola proclaimed that magic of this kind represented the culmination of all true philosophy [5].

"But the magus was not the only new figure of power to stalk the landscape of Europe in the fifteenth century. In Florence and Siena, Milan and Naples, engineers also flourished - engineers whom modern historians often identify, more one-sidedly than they would have identified themselves, as painters or architects." Grafton first introduces in his book Filippo Brunelleschi, then some less well-known men, like Mariano Taccola, from Siena, expert on hydraulics and military technology, who described his inventions in illustrated notebooks. There was also Georg Kyeser, a German military expert who did the same, even more resplendently. "These engineers were tasked by the communes and warlords who employed them, not only with practical jobs like making cannon and building fortifications, but also with what amounted to magical assignments ... From the fourteenth century on, they built the great escapement clocks that automatically tolled bells to signify the passing of the hours in cities and monasteries, and sometimes turned wheels to indicate the movement of the planets. Engineers equipped these timepieces with magnificent "jacks"- automata. [3]" In fact, the engineers performed a vital service for the rulers of cities, the service of organizing their feats when occasion required them. To show the power on nature and men of their lords, engineers built automata and spectacular pageant wagons, which moved apparently without anybody pulling them. "The engineers who carried out these feats - like their rivals, the magicians - drew on ancient technologies: not spells, in this case, but the hydraulic and pneumatic technology of ancient Alexandria, which had been transmitted through the Islamic and Byzantine worlds to the west. Engineers, like magicians, loved to amaze and frighten their audiences". [3]

**Giovanni de la Fontana**
And then, in [3] Giovanni Fontana, an early fifteenth-century engineer, is discussed. Fontana was the first to draw an illustration of a magic lantern, devised for the projection of devils in order to frighten people. "And he used his engineering skills to mimic the magicians' tours de force - as when, for example, he showed that by attaching articulated skeletons to a massive clockwork device, one could literally make the dead walk - exactly what necromancers claimed to do with their spells." Besides the creation of such devices, he applied his knowledge to analyze some natural phenomena, and to produce a treatise on perspective, which he presented to the painter Jacopo Bellini.

Let us report an important remark that Grafton writes about the Fontana's work. "Modern scholars often note that early engineers did not supply formal working drawings of their devices, but represented them in real time, functioning, in a way that did not give away their secrets but could appeal to patrons. Fontana, however, makes a superb exception to this rule. ... He drew not only male and female devils inspiring terror in real time by their fearsome appendages, but also the underlying mechanisms, which he laid out with the abstracting brilliance of a fifteenth-century Giacometti or Max Ernst. ... Fontana insisted that he was no magus. When witnesses at Padua exclaimed that a torpedo he had designed must run by diabolic power, he refuted them with contempt: the device was purely mechanical, as befitted a maker who was also a master of both medieval Archimedean statics and optics and of Renaissance engineering craft." [3]

Giovanni de la Fontana was born in Venice, probably in the last decade of the XIV century: it is supposed in the year 1390, based on the fact that he received a doctorate from the University of Padua in 1418 [6]. Very little is known about his childhood and his adolescence. He probably spent his youth in Venice or nearby this town, because he could remember many years later, in his "Liber de omnibus rebus naturalibus", the great fear of the town due to an awful phenomenon, and almost certainly it was the "great wind" of 1410 [6]. We can obtain some other data from the documents of the University of Padua: on 26 May 1417, designated as a "teacher", he witnessed some examinations. On June 18, 1418 in the bishop's palace, and on the 19th of the same month at the Cathedral, Fontana took the final exams to obtain the license and the recital of the public lecture for the doctorate. On 17 may 1421, he received his degree, becoming then a doctor in medicine, and appearing in the documents as "magister Iohannes Fontana de Venetiis artium doctor".

Between 1419 and 1440, Fontana served as doctors [6]. According to what he is telling in his "De trigono balistario", this activity was quite intense, almost impeding his favorite studies of engineering, hampered by the lack of libraries in those places too. However, Fontana took some advantage from his condition: in particular, being on the mountains, he experienced some trigonometric measurements with an instrument that he designed, which he described in a large treatise, apparently lost, and in the "De trigono balistario" [6]. It is unknown how long he spent in that place, but he was in Udine in 1440 for sure, when he ended this work.

Fontana spent the last period of his life in writing the "Liber de omnibus rebus naturalibus", written after the "De trigono balistario" of 1440. Probably, he composed his last book between the jubilee of 1450, because he reported it in the book, and year 1454. It seems that Fontana during this same period went to Rome and Crete. The date and place of Fontana's death is still unknown, however after 1454 [6].

**Fontana's works**

First Fontana was interested in the design of clocks, as evidenced by two treatises in a code of the University of Bologna. The first work is entitled "Nova compositio horologii", ended in 1418 in Padua, and dedicated to his friend Ludovicus Venetus, "cum studuit in artibus et medicina." According to Clagett [7], this is the first work of Fontana, and there he is declaring his intention to write a treatise entitled "Liber de ponderibus", a work on military devices, and, perhaps another treatise on the aesthetic appearance of machines [6]. A list of Fontana's works according to Clagett is given in the Appendix.

The second work is entitled "Horologium aqueum", discussing the same subjects of the first treatise. The author of the "Horologium aqueum" is referred to as "celeberrimus artium et medicinae doctor magister Iacobus Foritana de Venetiis" and announced the intention to write treatises entitled "De ponderibus", "Tractatus diversorum modorum horologii mixti", "De motibus aquarum" and the "Tractatus de rotalegis omnium generum", devoted to discuss the use of wheels in clocks, or, more generally, the use of wheels in continuous or perpetual motion. The intent of Fontana's first treatise, accompanied by good illustrations, was that of matching a clepsydra and some mechanics to obtain a smooth and constant movement, the first step towards the realization of a wheel in perpetual motion [8]. In the second treatise, the discussion focuses on the description of the water-clock and how it works, and it seems a plan for writing on various other types of instruments for measuring time using the movements of the four elements [6].

Another treatise written by Fontana, when he was young, is the "Tractatus/metrologum de pisce, cane et volucre", partially published by Thorndike in 1934 [6,1]. In this treatise, Fontana is combining some magic illusions with mechanical experiments. He describes some methods to measure together time and motion in space. Here, the author's aim is the design of clocks, not only using water and earth, but also air and fire, for instance, by measuring the time of the wax consumption of a burning candle or oil in lamps [6]. There is also the idea that it is possible to create some mechanical clocks able of measuring short time intervals through the movement of wheels, which make a complete turn over different periods: one hour, one minute, one second. These instruments were also proposed for the measurement of unknown heights or depths by throwing stones from the top of towers or into wells. The description of these experiments is accompanied by the narration of some episodes found in the ancient histories of Alexander or of other men, who were able to fly or dive into the sea abysses [6].

These inventions and measurements were often accompanied by some magic illusions: Fontana built some shapes appearing as evils, filled with chemical substances able to cause their movement under water and the emission of rays, as he said, these shapes terrified a too curious monk [6]. In Padua, he claims to have made other deceptive tricks that caused some learned and sagacious men to state that he had summoned from hell demonic spirits using necromantic arts, with the exorcism and other occult sciences.

In the "Tractatus de pisce" are also reminded the first treatises on clocks and he is also using the theory of impetus. Here we can find an expression of interest for the military art that will be displayed in the "Bellicorum instrumentorum liber cum figuris et fictitiis literis conscriptus", an encrypted code (cod. Icon. 2112 of the Bayerische Staatsbibliothek, Munchen, [9]), however, accompanied by headings and introductory notes in plane writing. The work does not contain any indication of the date of its composition, but scholars are unanimously considering it composed after three written texts in the code of Bologna, between 1420 and 1440.

"Bellicorum instrumentorum liber" is a notebook which describes a large number of tools and machines, with illustrations and short explanations, and goes far beyond the subject indicated in the title. We find several military engines, such as weapons and rams in a siege engine, but also devices for fountains, tricks to deceive the senses as in a "castle of deception", a maze, some special stage effects such as the shining miter and crozier and a magic candlestick [6]. He illustrates the designs of musical instruments such as mechanical organs (Puerilia) and masks, keys and locks, warships, double mirrors, stoves and surgical instruments. The code contains a part dedicated to hydraulic projects, public fountains, systems of water distribution, experiments with siphons, and also alembics and alchemical vessels for two or more liquids. According to [10,6], this manuscript shows that Fontana aimed to investigate a series of devices from ancient books of mathematicians and naturalists such as Archimedes, Heron, Philo, and even Ovid and Pliny the Elder, and the Arab writers, mainly from Al-Kindi, to the his contemporary artists and craftsmen.

Probably, during the years when Fontana met the Count of Carmagnola (1428-1432 c.), he prepared the "Secretum de thesauro experimentorum ymaginationis hominum", in encripted writing, except for the titles of chapters. In this work, Fontana is investigating the different types of memory, and explains the functions of the artificial memory [11]; he describes also several mnemonic techniques. Fontana considered the artificial memory a necessary tool and of common use such as reading and writing. Here Fontana uses Aristotle and other philosophy's doctors, however, he is adding to the natural memory the ability to produce a set of representations that can be manipulated and combined according to the will and cognitive purposes of the user. Fontana proposes some memory devices and "machines", having a fixed structure (wheels, spirals, cylinders) and a mobile and variable part allowing to change the combinations of signs within this system.

The work provides several mnemonic tricks too and a rich iconographic repertoire: the text is ciphered using a secret alphabet established on the basis of a symbolic view of signs, and therefore used not to conceal the text but to enhance the mnemonic mechanism [10].

The "De trigono balistario" was ended the last day of February 1440 in Udine. In the work, partially published by Clagett in his "Archimedes in the Middle Ages" ([7], pp. 270-294) there is described the sextant, its construction and how to use it. This Fontana's sextant was a trigonometric instrument, designed and developed to have from an optical measurement some data that could be reported on some maps. His aim was that of having a more reliable geographical collection of data, to correct wrong longitudes and latitudes. On optics, we have to remember the notes written by Fontana to the Alhazen's "Liber de speculis comburentibus", where he is discussing de sectione mukesi (or mukefi, mukafi, parabola), that is, of the parabolic section.

The "Liber de omnibus rebus naturalibus quae continentur in mundo, videlicet coelestibus et terrestribus nec non mathematicis et de angelis motoribus quae coelorum" was published in Venice in 1544, under the name of Pompilio Azali. The author was recognized in Fontana by Thorndike [1,6]. This work was composed in 1454 and it is a summa of his encyclopedic knowledge, subdivided in five sections, about "quasi de omnibus quae sunt in mundo", on visible and invisible things, with theological and philosophical questions related to them. We find the discussion of the orbits of the planets and stars with their motions and properties, the four elements that form the objects of experience, and then the regions of the Earth with all you can find in them, such as animals, plants and minerals. At the beginning of the book, Fontana rejects the hypothesis of the eternity of the world, attributing it to Aristotle, because this is in contrast with the faith. Moreover, he claims that the celestial influences do not overwhelm the human will, which remains in some amount free [6]. The description of the universe in its various parts allows Fontana showing his vast knowledge about astronomy and geography.

**The Bellicorum instrumentorum**
From link at Ref.9, we can see the pages of Fontana's work entitled "Bellicorum instrumentorum liber, cum figuris et fictitys litoris conscriptus", that is, Illustrated and encrypted book of war instruments. As previously told, we find in it a wide range of technologies, and not only war machines, included in 70 folio pages with some 140 illustrations. "Fontana's Bellicorum instrumentorum features, among others, siege engines, fountains and pumps, lifting and transporting machines, defensive towers, dredges, combination locks, battering rams, rocket-propelled animals, the first ever depiction of a magic lantern, scaling ladders, measuring instruments, alchemical furnaces and … several robotic automata" [12]. Almost each figure is accompanied by a short text. The first sentence is usually in Latin, with the remainder written in cypher. "There was something of a magician in Fontana, … who is thought to have built toy models of a few of the automata, so projecting a veil of esoteric mysticism will have appealed to his desire to amaze readers" [12]. According to Ref.12, the coded writing may also have offered a form of copyright, because any unauthorized copying was difficult.

Figure 1 shows the "first ever depiction of a magic lantern", described as "apparentia nocturna nel tarore videntius", using the image of a devil. The rocket-propelled device, show in Figure 2, was tested by Richard Windley [13,14], and it works. Among the illustrations, there are some sailing castles, that is, large structures propelled by wind (Figure 3).

I personally find interesting the machine in the Figure 4, a dredge. The dredges were used to build waterways. The Fontana dredge is installed on a pontoon and its working tool was a dipper moving in a chute with a sharp metal tip. The first record of the design of such a device is in fact that of Fontana [15]. Besides the weapons and the devices able to create Fontana's "artificial magic", we find then in his book several examples of technologies of the early modern age in Europe.

**Appendix**
List of the works by Giovanni Fontana, from the web page: www.voynich. net/neal/fontana_works.html, abstracted from Marshall Clagett, The Life and Works of Giovanni Fontana, [7].
- Nova compositio horologii, Bologna, Bibl. Univ. 2705, 1r-51v. 15th century.
- Horologium aqueum, Bologna, Bibl. Univ. 2705, 1r-51v. 15th century.
- Tractatus (or Metrologum) de pisce, cane et volucre, Bologna, Bibl. Univ.,15th century.
- Bellicorum instrumentorum liber cum figuris et fictitiis literis conscriptus, Munich, Staatsbibl.
- Secretum de thesauro experimentorum ymaginationis hominum, Paris.
- Notes on the Liber de speculis comburentibus of Alhazen,Paris.
- Tractatus de trigono balisterio, Oxford.
- Liber de omnibus rebus naturalibus quae continentur in mundo, videlicet coelestibus et terrestribus nec non mathematicis et de angelis motoribus quae coelorum, Venice, 1544.
- A medical recipe of Magister Johannes Font. de Venetiis, MS no 101 (147) of the library of Baldassarre Boncompagni at Rome.
Works not yet discovered:
- Liber de ponderibus.
- Libellus de aquae ductibus.
- De laberintis libellus.
- Artis pictoriae canones ad Iacobum Bellinum.
- De spera solida.
- Tractatus major de trigono balistario.
- De speculo medicinali.
- De rotalegis omnium generum.
- Tractatus diversorum modorum horologii mixti.
Of doubtful attribution:
- De speculo mukefi, Oxford, Bodl. Libr. Canon. Misc. 480, 47r-54r. 16th century, Florence, Bibl. Med. Laur. Ashburn. 957, 95r-110v. 15th-16th century; Vienna, Nationalbibl. 5258, 27r-38v. 15th century; Verona, Bibl. Capitolare 206, 1r-8v. 16th century; London, British Museum Cotton Tib. B. IX, 231r-35v. 15th century.
- Protheus, Florence, Bibl. Med. Laur. Ashburn. 957, 71r-94v. Late 15th or early 16th century.

**References**
1. L. Thorndike, A history of magic and experimental science, New York-London, Columbia University Press, 1923-58.
2. W. Royall Newman and A. Grafton Secrets of Nature: Astrology and Alchemy in Early Modern Europe, , MIT Press, 2001 -
3. A. Grafton, Magic and technology in early modern Europe, published by the Smithsonian Institution Libraries, Washington, DC, 2002
4. D. Stolzenberg The Great Art of Knowing: The Baroque Encyclopedia of Athanasius Kircher, Stanford University Libraries, Small Press Distribution, 2001.
5. Giovanni Pico della Mirandola, Oratio de hominis dignitate; Oration on the dignity of man; English translation by Elizabeth Livermore Forbes (Anvil Press, 1953).
6. Maria Muccillo, Giovanni (de Fontana, de la Fontana), Dizionario Biografico degli Italiani - Volume 48 (1997), available at http://www.treccani.it/enciclopedia/fontana-giovanni-antonio-jacopo_(Dizionario-Biografico)/
7. Marshall Clagett, The Life and Works of Giovanni Fontana. Annali dell'Istituto e museo di storia della scienza di Firenze 1 (1976), pp.5-28; Archimedes in the Middle Ages, University of Wisconsin Press. In 5 volumes, texts in Latin and English.

8. Derek J. de Solla Price, On the Origin of Clockwork, Perpetual Motion Devices, and the Compass, United States National Museum, Bulletin 218, Contributions from The Museum of History and Technology, Smithsonian Institution, Washington, DC, 1959.
9. Giovanni Fontana, Bellicorum instrumentorum liber cum figuris et fictitiis literis conscriptus, cod. Icon. 2112, Bayerische Staatsbibliothek, Munchen. http://codicon.digitale-sammlungen.de/bsb00001353Cod.icon.%20242%20a.html
10. E. Battisti and G. Saccaro Battisti. Le Macchine Cifrate Di Giovanni Fontana: Con La Riproduzione Del Cod. Icon. 242 Della Bayerische Staatsbibliothek Di Monaco Di Baviera E La Decrittazione Di Esso E Del Cod. Lat. Nouv. Acq. 635 Della Bibliothèque Nationale Di Parigi. Milano: Arcadia, 1984.
11. http://en.wikipedia.org/wiki/Art_of_memory
12. http://history-computer.com/Dreamers/Fontana.html
13. Richard Windley, Technical and historical reproduction for film and television, http://www.richardwindley.co.uk/
14. http://www.richardwindley.co.uk/fontana-rocket-car.html
15. K-H. Grote, E.K. Antonsson, Springer Handbook of Mechanical Engineering, Volume 10, Springer, 2009

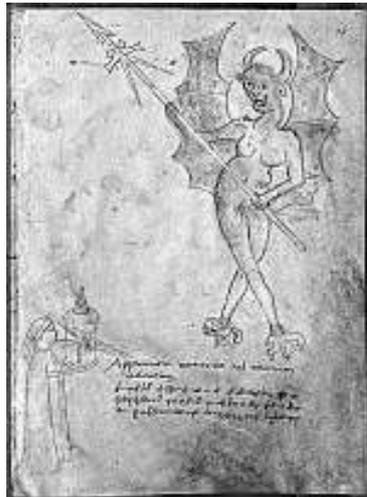
Fig.1. The magic lantern

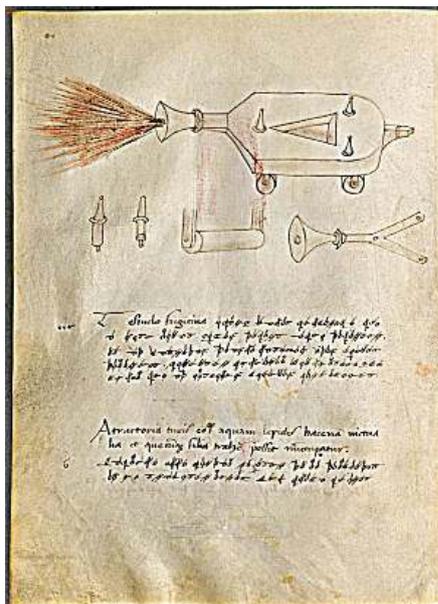
Fig.2. The rocket engine.

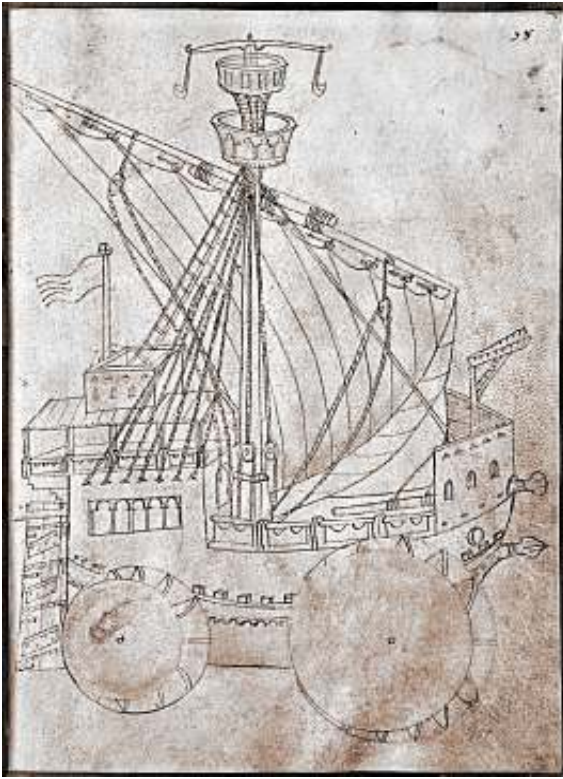
Fig.3 The sailing castle.

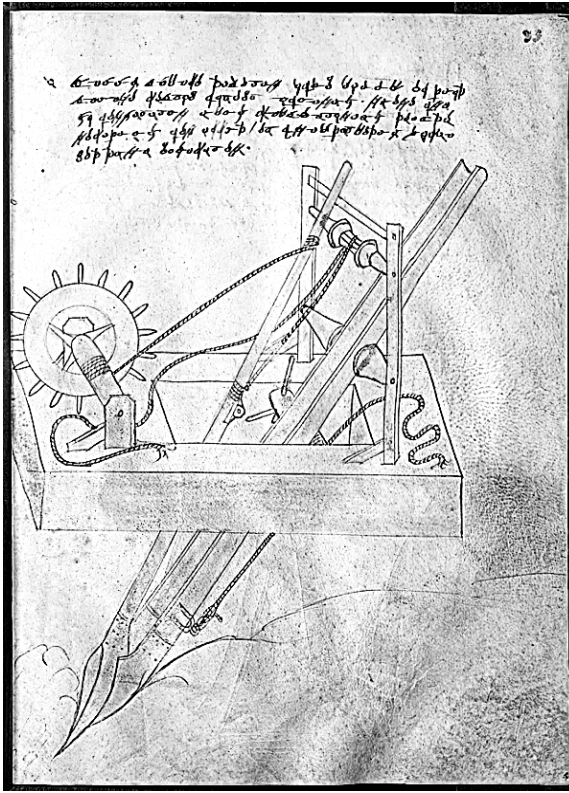
Fig.4 Fontana's dredge.